%% file: HigherOrder.tex
\documentclass[a4paper,conference]{IEEEtran}
\usepackage{etex}
\usepackage{graphicx,ifthen}
\usepackage{psfrag,amssymb,amsthm}
\usepackage[cmex10]{amsmath}

\usepackage{mathtools}
\usepackage[noadjust]{cite}

\usepackage{amsmath}
\usepackage{amsthm}
\usepackage{amssymb}
\usepackage{bm}
\usepackage{xspace}
\usepackage{xcolor}
\usepackage{url}
\usepackage{framed}
\usepackage{float}
\usepackage{rotating}
\usepackage{verbatim}
\usepackage{listings}
\usepackage{lscape}
\usepackage{ifthen}
\usepackage{pstricks,pst-sigsys}
\usepackage{subfigure}

 \usepackage{pgfplots}
 \usepackage{pgf}
 \usepackage{tikz}
 \usetikzlibrary{arrows,shapes.misc,chains,scopes,fit}
 \pgfplotsset{compat = newest}
 \usepackage{pgfplotstable}

 \usepackage{booktabs}
 \usepackage{colortbl}

\newtheorem{lem}{Lemma}

\theoremstyle{definition}
\newtheorem{definition}{Definition}

\newtheorem{example}{Example}

\title{Higher-Order Kullback-Leibler Aggregation of Markov Chains}

\author{
\IEEEauthorblockN{Bernhard C. Geiger, Yuchen Wu}\\
\IEEEauthorblockA{Institute for Communications Engineering, Technical University of Munich, Germany\\
geiger@ieee.org}
}

\input{abbrevations_processing.tex}

\newcommand{\Qvec}{\mathbf{Q}}
\newcommand{\Emat}{\mathbf{E}}

\newcommand{\blocked}[2]{{#1}^{(#2)}}
\newcommand{\Marx}[2]{\tilde{#1}^{M(#2)}}
\renewcommand{\entrate}[1]{\bar{H}(#1)}
\renewcommand{\redrate}[1]{\bar{R}(#1)}

\newcommand{\identity}{\mathrm{id}}

\begin{document}
\maketitle

\begin{abstract}
In this work, we consider the problem of reducing a first-order Markov chain on a large alphabet to a higher-order Markov chain on a small alphabet. We present information-theoretic cost functions that are related to predictability and lumpability, show relations between these cost functions, and discuss heuristics to minimize them. Our experiments suggest that the generalization to higher orders is useful for model reduction in reliability analysis and natural language processing.
\end{abstract}


\section{Introduction}\label{sec:intro}
In many scientific disciplines, the Markov chains used to model real-world stochastic processes are too large to admit efficient simulation or estimation of model parameters. For example, in natural language processing and computational chemistry, the size of the Markov chain's alphabet is related the number of words in a dictionary and the number of molecules in a given volume, respectively. A popular technique to deal with the problem is \emph{aggregation} (see Fig.~\ref{fig:aggregation} and Definition~\ref{def:problem} below), i.e., partitioning the alphabet and approximating the (generally non-Markovian) process on this partitioned alphabet by a first-order Markov chain on this partition.

Solving the aggregation problem thus requires finding an appropriate partition of the alphabet, or equivalently, a non-injective function from the alphabet to a smaller set. Given the cardinality of this smaller set, Vidyasagar~\cite{Vidyasagar_MarkovAgg} and Deng et al.~\cite{Meyn_MarkovAggregation} proposed an information-theoretic quantity as a cost function for solving this problem. Deng et al.\ argued that choosing a partition function that minimizes this cost maximizes the predictability of the aggregation, i.e., the future state of the aggregation can be inferred from the current state with high probability. This is particularly useful if the original Markov chain models a process in which groups of states behave almost deterministically (e.g., the process oscillates between groups of states or stays within each group for a long time). For nearly completely decomposable Markov chains, the authors of~\cite{Meyn_MarkovAggregation} found a connection between their information-theoretic cost function and spectral, i.e., eigenvector-based aggregation techniques. 

Based on~\cite{Meyn_MarkovAggregation}, the authors of~\cite{GeigerEtAl_OptimalMarkovAggregation} proposed another information-theoretic cost function, the minimization of which is equivalent to making the process on the partitioned alphabet ``as Markov as possible''. Their analysis is based on a recent information-theoretic characterization of lumpability~\cite{GeigerTemmel_kLump}, the scenario in which the process on the partitioned alphabet has the Markov property. The authors of~\cite{GeigerEtAl_OptimalMarkovAggregation} showed that, in this formulation, the aggregation problem can be sub-optimally solved by applying the information bottleneck method~\cite{Tishby_InformationBottleneck}, a popular method in machine learning.

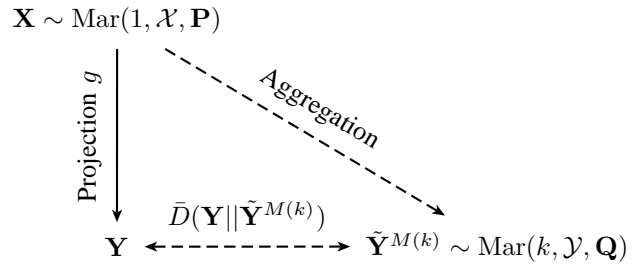
\begin{figure}[t] 
\centering
  \begin{pspicture}[showgrid=false](-1,1)(6,4.25)
    \psset{style=Arrow}
    \pssignal(0,4){x}{$\Xvec\sim\Mar{1,\dom{X},\Pvec}$}
    \pssignal(0,1){y}{$\Yvec$}
    \pssignal(5,1){yp}{$\Marx{\Yvec}{k}\sim\Mar{k,\dom{Y},\Qvec}$}
    \ncline{<-}{y}{x} \aput{:U}{Projection $g$}
    \ncline[style=Dash]{x}{yp}\aput{:U}{Aggregation}
    \pnode(4.8,1.4){yp1}\pnode(4.8,4.6){xp1}\pnode(5.2,1.4){yp2}\pnode(5.2,4.6){xp2}
    \ncline[style=Dash]{<->}{y}{yp}\Aput{$\kldr{\Yvec}{\Marx{\Yvec}{k}}$}
\end{pspicture}
\caption{Illustration of the aggregation problem: A Markov chain $\Xvec$ is given. We are interested in finding a function $g{:}\ \dom{X}\to\dom{Y}$ and an aggregation of $\Xvec$, i.e., a $k$-th order Markov chain $\Marx{\Yvec}{k}$ on $\dom{Y}$. The function $g$ defines a process $\Yvec$ via $Y_{g,n}:=g(X_n)$, the projection of $\Xvec$. $\Yvec$ might not be Markov of any order, but can be approximated by a $k$-th order Markov chain $\Marx{\Yvec}{k}$.}
\label{fig:aggregation}
\end{figure}

In this work, we generalize these two approaches to \emph{higher-order} aggregations, i.e., propose information-theoretic cost functions for approximating the process on the partitioned alphabet by a $k$-th order Markov chain. We show that extending the results of~\cite{GeigerEtAl_OptimalMarkovAggregation} relates to higher-order lumpability (Section~\ref{sec:Aggregation:lump}), while extending the results of~\cite{Meyn_MarkovAggregation} amounts to making the future state of the aggregation predictable from the last $k$ states (Section~\ref{sec:Aggregation:pred}). In Sections~\ref{sec:Aggregation:interplay} and~\ref{sec:Aggregation:lift} we discuss the properties of the corresponding cost functions and show that they form an ordered set. After briefly discussing heuristic algorithms for minimizing these cost functions in Section~\ref{sec:Aggregation:algos}, Section~\ref{sec:Applications} illustrates our findings on synthetic and real-world models.

\section{Notation and Definitions}\label{sec:prelim}
All random variables (RVs) and stochastic processes are defined on the probability space $(\Omega,\mathfrak{B},\mathrm{Pr})$. An RV $Z$ takes values $z$ from a finite set $\dom{Z}$. The probability mass function (PMF) of $Z$ is denoted by $p_Z$, where $p_Z(z) := \Prob{Z=z}$ for all $z\in\dom{Z}$. Joint and conditional PMFs are defined similarly. 

A discrete-time stochastic process $\Zvec$ on $\dom{Z}$ is a one-sided sequence of RVs $(Z_1,Z_2,\dots)$, each RV taking values from $\dom{Z}$. We abbreviate $Z_m^n:=(Z_m,Z_{m+1},Z_n)$. The processes considered in this work are \emph{stationary}, i.e., PMFs are invariant w.r.t.\ a time shift. In particular, the marginal distribution of $Z_k$ is equal for all $k$ and shall be denoted as $p_Z$.

We will use information-theoretic quantities as cost functions for aggregation. Specifically, let $\ent{Z}$ be the entropy of $Z$, $\ent{Z_2|Z_1}$ the conditional entropy of $Z_2$ given $Z_1$, and $\mutinf{Z_1;Z_2}:=\ent{Z_2}-\ent{Z_2|Z_1}$ the mutual information between $Z_1$ and $Z_2$~\cite[Ch.~2]{Cover_Information2}. Furthermore, the entropy rate and the redundancy rate of a stationary stochastic process $\Zvec$ are
\begin{subequations}
\begin{align}
 \entrate{\Zvec}&:= \limn \ent{Z_n|Z_{1}^{n-1}}\\
 \redrate{\Zvec}&:= \limn \mutinf{Z_n;Z_1^{n-1}} = \ent{Z}-\entrate{\Zvec}.\label{eq:redrate}
\end{align}
\end{subequations}
The redundancy rate measures how much information the past and current states of a process contain about its future state. In some sense, the redundancy rate measures the predictability of a stochastic process. Indeed, for a deterministic process $\Zvec$ we have $\entrate{\Zvec}=0$ and the redundancy rate achieves its maximum at $\redrate{\Zvec}=\ent{Z}$. 

\begin{definition}[Kullback-Leibler Divergence Rate]\label{def:KLDR}
 The Kullback-Leibler divergence rate (KLDR) between two stationary stochastic processes $\Zvec$ and $\Wvec$ on the same finite alphabet $\dom{Z}$ is~\cite[Ch.~10]{Gray_Entropy}
\begin{align}\label{eq:KLDRdef}
 \kldrate{Z}{W}&:= \limn \frac{1}{n}\sum_{z_1^n\in\dom{Z}^n} p_{Z_1^n}(z_1^n) \log \frac{p_{Z_1^n}(z_1^n)}{p_{{W}_1^n}(z_1^n)}
\end{align}
whenever the limit exists and if, for all $n$ and all $z_1^n$, $p_{{W}_1^n}(z_1^n) = 0$ implies $p_{Z_1^n}(z_1^n) = 0$ (short: $p_{Z_1^n}\ll p_{{W}_1^n}$).
\end{definition}

The KLDR can be used to measure the (dis-)similarity between two stochastic processes. Gray discussed examples for which the limit in~\eqref{eq:KLDRdef} exists~\cite{Gray_Entropy}, and Rached et al.\ evaluated the limit between Markov chains (not necessarily stationary or irreducible)~\cite{Rached_KLDR}.

\section{Markov Chains}\label{sec:Markov}
Let $\Zvec$ be an irreducible and aperiodic, time-homogeneous, $k$-th order Markov chain on the alphabet $\dom{Z}=\{1,\dots,N\}$ with a $(k+1)$-dimensional transition matrix $\Pvec=\{P_{i_1,i_2,\dots,i_k\to j}\}$. Here, $ P_{i_1,i_2,\dots,i_k\to j} := p_{Z_n|Z_{n-k}^{n-1}}(j|i_1,i_2,\dots,i_k)$, i.e., $\Pvec$ is row stochastic. If the $k$-dimensional initial distribution $p_{Z_1^k}$ is invariant under $\Pvec$, i.e., if, for all $i_1,\dots, i_{k+1}\in\dom{Z}$,
\begin{equation*}
 p_{Z_1^k}(i_2,\dots,i_{k+1}) = \sum_{i_1\in\dom{Z}}p_{Z_1^k}(i_1,\dots,i_k) P_{i_1,i_2,\dots,i_k\to i_{k+1}}
\end{equation*}
then $\Zvec$ is \emph{stationary}. For irreducible Markov chains, this invariant distribution is unique and we use the shorthand notation $\Zvec\sim\Mar{k,\dom{Z},\Pvec}$. In particular, if $k=1$ and $\mu_i:=p_Z(i)$, we have $\muvec^T=\muvec^T\Pvec$~\cite[Thm.~4.1.6]{Kemeny_FMC}.

\begin{lem}[{\cite[Prop.~3]{GeigerTemmel_kLump}}]\label{lem:ITMarkov}
 $\Zvec\sim\Mar{k,\dom{Z},\Pvec}$ if and only if $\entrate{\Zvec}=\ent{Z_{k+1}|Z_1^k}$.
\end{lem}
This information-theoretic characterization can be used to show that the \emph{$k$-transition chain} $\blocked{\Zvec}{k}$ of a $k$-th order Markov chain with states $\blocked{Z}{k}_1=Z_1^k$, $\blocked{Z}{k}_2=Z_{2}^{k+1}$,\dots\ is a first-order Markov chain on the alphabet $\dom{Z}^k$ having a transition matrix
\begin{multline}\label{eq:TransitionChain}
 \tilde{P}_{(i_1,i_2,\dots,i_k)\to (j_1,j_2,\dots,j_k)}=  \\
 \begin{cases}
  P_{i_1,i_2,\dots,i_k\to j_k}, & (i_2,\dots,i_k)=(j_1,j_2,\dots,j_{k-1})\\ 0, &\text{else}
 \end{cases}.
\end{multline}
The KLDR between two $k$-th order Markov chains $\Zvec\sim\Mar{k,\dom{Z},\Pvec}$ and $\Zvec'\sim\Mar{k,\dom{Z},\Pvec'}$ equals
\begin{multline}\label{eq:KLDRMarkov}
 \kldrate{Z}{Z'}=\\
  \sum_{i_1,\dots,i_k,j\in\dom{Z}} p_{Z_1^k}(i_1,\dots,i_k)P_{i_1,i_2,\dots,i_k\to j} \log\frac{P_{i_1,i_2,\dots,i_k\to j}}{P'_{i_1,i_2,\dots,i_k\to j}}
\end{multline}
where $p_{Z_1^k}$ and $p_{{Z'}_1^k}$ are invariant under $\Pvec$ and $\Pvec'$, and where it is assumed that $P'_{i_1,i_2,\dots,i_k\to j}=0$ implies $P_{i_1,i_2,\dots,i_k\to j}=0$ and that $p_{Z_1^k}\ll p_{{Z'}_1^k}$. Note that for $k=1$,~\eqref{eq:KLDRMarkov} simplifies to~\cite{Rached_KLDR}
\begin{align}
\kldrate{Z}{Z'}&= \sum_{i,j\in\dom{Z}} \mu_i P_{i\to j}\log\frac{P_{i\to j}}{P'_{i\to j}}. \label{eq:KLDRMarkov1}
\end{align}

\section{Aggregating Markov Chains}\label{sec:Aggregation}
Let $\Xvec\sim\Mar{1,\dom{X},\Pvec}$ be a first-order Markov chain with alphabet $\dom{X}=\{1,\dots,N\}$ and consider a non-injective function $g{:}\ \dom{X}\to\{1,\dots,M\}=:\dom{Y}$, $M<N$. Note that $g$ induces a partition of $\dom{X}$. Since $\Xvec$ is stationary, we can define a stationary process $\Yvec$ with samples $Y_n:=g(X_n)$. We call $\Yvec$ the \emph{projected process}, or simply the \emph{projection}, cf.~Fig.~\ref{fig:aggregation}. The data processing inequality~\cite[Ch.~2.8]{Cover_Information2} implies that $\ent{Y}\le\ent{X}$, $\entrate{\Yvec}\le\entrate{\Xvec}$, and $\redrate{\Yvec}\le\redrate{\Xvec}$.

In general, the projection $\Yvec$ is not Markov of any order. Nevertheless, it is possible to approximate $\Yvec$ by a Markov chain $\Marx{\Yvec}{k}\sim\Mar{k,\dom{Y},\Qvec}$ that is close to $\Yvec$ in the sense of minimizing $\kldr{\Yvec}{\Marx{\Yvec}{k}}$, cf.~Fig.~\ref{fig:aggregation}. The transition matrix $\Qvec$ minimizing this KLDR satisfies~\cite[Cor.~10.4]{Gray_Entropy}
\begin{equation}\label{eq:QfromX}
 Q_{i_1,i_2,\dots,i_k\to j} = p_{Y_{k+1}|Y_1^k}(j|i_1,i_2,\dots,i_k)
\end{equation}
and we get
\begin{equation}
 \kldr{\Yvec}{\Marx{\Yvec}{k}} = \ent{Y_{k+1}|Y_1^k}-\entrate{\Yvec}.
\end{equation}

\begin{definition}[Markov Aggregation]\label{def:problem}
Let $\Xvec\sim\Mar{1,\dom{X},\Pvec}$ and $\dom{Y}$ be given. The problem of Markov aggregation concerns finding a function $g{:}\ \dom{X}\to\dom{Y}$ and a $k$-th order Markov chain $\Marx{\Yvec}{k}\sim\Mar{k,\dom{Y},\Qvec}$, both of which are optimal w.r.t.\ a well-defined cost function.
\end{definition}

In what follows we will concretize this very loosely stated definition. In particular, as we showed above, for a given function $g$, the $k$-th order Markov chain $\Marx{\Yvec}{k}\sim\Mar{k,\dom{Y},\Qvec}$ minimizing the KLDR to the projection $\Yvec$ is given by~\eqref{eq:QfromX}. It remains to find a ``good'' function $g$. We will introduce two different approaches to finding such a $g$: The first approach tries to make $\Yvec$ as close as possible to a $k$-th order Markov chain, while the second approach tries to make $\Marx{\Yvec}{k}$ as predictable as possible. Both approaches can be described by information-theoretic cost functions.

\subsection{Aggregation via Lumpability}\label{sec:Aggregation:lump}
Let us start with the goal to find $g$ such that $\Yvec$ is close to a $k$-th order Markov chain. The rare scenarios in which $\Yvec$ \emph{is} a $k$-th order Markov chain are categorized with the term \emph{lumpability}:

\begin{definition}[Lumpability{~\cite[Sec.~6.3]{Kemeny_FMC}}]\label{def:lump}
 A Markov chain $\Xvec\sim\Mar{1,\dom{X},\Pvec}$ is \emph{$k$-lumpable} w.r.t.\ a function $g$, iff the projection $\Yvec$ is $\Mar{k,\dom{Y},\Qvec}$ for every initial distribution $p_{X_1}$ of $\Xvec$, and if $\Qvec$, given by~\eqref{eq:QfromX}, does not depend on this initial distribution.
\end{definition}

Conditions for 1-lumpability have been presented in~\cite[Thm.~6.3.2]{Kemeny_FMC}; for a general $k$, a linear-algebraic condition can be found in~\cite{GurvitsLedoux_MarkovPropertyLinearAlgebraApproach}. The following characterization is information-theoretic:

\begin{lem}[{\cite[Thm.~2]{GeigerTemmel_kLump}}]\label{lem:ITLump}
 A Markov chain $\Xvec\sim\Mar{1,\dom{X},\Pvec}$ is $k$-lumpable w.r.t.\ $g$ if and only if
 \begin{equation}\label{eq:LumpCost}
  \Delta^k_L(\Xvec,g):=\ent{Y_{k+1}|Y_1^k} - \ent{Y_{k+1}|Y_2^k,X_1} =0.
 \end{equation}
\end{lem}
The authors of~\cite{GeigerEtAl_OptimalMarkovAggregation} used $\Delta^1_L(\Xvec,g)$ as a cost function for Markov aggregation. Their goal was, as ours in this section, to find a partition of the original alphabet $\dom{X}$ w.r.t.\ which the Markov chain $\Xvec$ is ``most 1-lumpable''. The definition of $\Delta^k_L(\Xvec,g)$ now provides a cost function for higher-order aggregation.

Consider the two extreme cases $M=N$ and $M=1$; the first corresponds to the identity function $g=\identity$, the second to the constant function $g\equiv 1$. In both cases, $\Yvec$ is a $k$-th order Markov chain, either because it coincides with $\Xvec$ or because it is constant. Hence, for every $k$, $\Delta^k_L(\Xvec,\identity)=\Delta^k_L(\Xvec,1)=0$. In contrast, there are clearly Markov chains that are not $k$-lumpable for any partition of $\dom{X}$ into $1<M<N$ elements. It immediately follows that neither $\Delta^k_L(\Xvec,g)$ is monotonous w.r.t.\ a refinement of the partition induced by $g$, nor that $\min_{g{:}\ \dom{X}\to\dom{Y}} \Delta^k_L(\Xvec,g)$ is monotonous w.r.t.\ the cardinality $M$ of $\dom{Y}$.

\subsection{Aggregation via Predictability}\label{sec:Aggregation:pred}
The authors of~\cite{Meyn_MarkovAggregation} focused on finding a partition w.r.t.\ which $\Xvec$ is nearly completely decomposable. In other words, for a well-chosen partition, transitions between different groups of states occur only with small probabilities, while transitions within such a group occur with high probabilities. The authors proposed to measure the ``decomposability'' of $\Xvec$ with
\begin{equation}\label{eq:DecCost}
 \Delta_D(\Xvec,g):=\mutinf{X_2;X_1} - \mutinf{Y_2;Y_1}.
\end{equation}
They observed that $\Delta_D(\Xvec,g)$ encompasses more than just near decomposability: Minimizing $\Delta_D(\Xvec,g)$ is related to making $\Yvec$ \emph{predictable}:

\begin{definition}[Predictability]
We call $\Yvec$ \emph{$k$-predictable}, iff there exists a predictor $r{:}\ \dom{Y}^k\to\dom{Y}$ such that
\begin{equation}
 \perr:=\Prob{r(Y_{n-k+1}^{n})\neq Y_{n+1}}=0.
\end{equation}
If $\Yvec$ is the projection of $\Xvec\sim\Mar{1,\dom{X},\Pvec}$, we call $\Xvec$ \emph{$k$-predictable w.r.t.\ $g$}.
\end{definition}

In~\cite{Verdu_GeneralizingFano}, the authors connected mutual information with prediction error probabilities and thus extended Fano's inequality~\cite[Ch.~2.10]{Cover_Information2}.  Specifically,~\cite[Thm.~5]{Verdu_GeneralizingFano} yields
\begin{align*}
 \mutinf{Y_{k+1};Y_1^{k}} &\ge \mutinf{Y_{k+1};r(Y_{1}^{k})} \\ &\ge -\log\left(\max_{y\in\dom{Y}}p_Y(y)\right)(1-\perr)-\binent{\perr}
\end{align*}
where $\binent{p}:=-p\log p - (1-p)\log(1-p)$ and where the first inequality is due to data processing~\cite[Ch.~2.8]{Cover_Information2}. Hence, minimizing $\Delta_P^k(\Xvec,g)$ admits a predictor $r$ with a small prediction error probability.

Based on this, we generalize~\eqref{eq:DecCost} to higher-order predictability and suggest the following family of cost functions for higher-order aggregation:
\begin{equation} \label{eq:PredCost}
 \Delta_P^k(\Xvec,g) := \mutinf{X_2;X_1} - \mutinf{Y_{k+1};Y_1^{k}}
\end{equation}
Obviously, $\Delta_P^1(\Xvec,g)\equiv\Delta_D(\Xvec,g)$.

In contrast to $\Delta^k_L(\Xvec,g)$, the data processing inequality shows that $\Delta^k_P(\Xvec,g)$ is monotonically decreasing w.r.t.\ a refinement of the partition, hence $\min_{g{:}\ \dom{X}\to\dom{Y}} \Delta^k_P(\Xvec,g)$ decreases monotonically with the cardinality $M$ of $\dom{Y}$. Nevertheless, even if a Markov chain $\Xvec$ is $k$-predictable w.r.t.\ $g$, it neither need to be $k$-predictable w.r.t.\ a refinement or a coarsening of $g$ (see the example in Section~\ref{ssec:ex:pred}). Considering again the two trivial cases $M=N$ and $M=1$ reveals another fact: For $M=N$ we have $\Delta^k_P(\Xvec,\identity)=0$, despite $\Yvec\equiv\Xvec$ not necessarily being $k$-predictable. For $M=1$ we have $\Delta^k_P(\Xvec,1)=\redrate{\Xvec}$, despite $\Yvec$ being $k$-predictable for every $k$. This apparent discrepancy results from adding the redundancy rate $\redrate{\Xvec}$ in~\eqref{eq:PredCost}, a quantity independent of $g$. Hence, while $\Delta^k_P(\Xvec,g)$ does not precisely quantify $k$-predictability, minimizing $\Delta^k_P(\Xvec,g)$ allows $\Yvec$ to be as $k$-predictable as possible. Note that this discrepancy is not present for lumpability, where  $\Delta_L^k(\Xvec,g)=0$ if and only if $\Xvec$ is $k$-lumpable w.r.t.\ $g$ (cf.~Lemma~\ref{lem:ITLump}).

\subsection{Interplay between Lumpability and Predictability}\label{sec:Aggregation:interplay}
The following example shows that 1-lumpability generalizes 1-predictability, which in turn generalizes decomposability.

\begin{example}\label{ex:blockstochastic}
Let $\Xvec\sim\Mar{1,\dom{X},\Pvec'}$, where
\begin{equation}\label{eq:blockstochastic}
 \Pvec' = \left[
 \begin{array}{cccc}
  a_{11} \Pvec_{11} & a_{12}\Pvec_{12}  &\cdots & a_{1M}\Pvec_{1M}\\
  a_{21} \Pvec_{21} & a_{22}\Pvec_{22}  &\cdots & a_{2M}\Pvec_{2M}\\
  \vdots & \vdots & \ddots & \vdots\\
  a_{M1} \Pvec_{M1} & a_{M2}\Pvec_{22}  &\cdots & a_{MM}\Pvec_{MM}
\end{array}
 \right].
\end{equation}
Here $\Amat=\{a_{ij}\}$ and all $\Pvec_{ij}$ are row stochastic matrices. Moreover, the diagonal blocks $\Pvec_{ii}$ are square with size $N_i \times N_i$. Assume that $g$ induces the same partition that can be derived from $\Pvec'$, i.e., $\dom{X}=\{1,\dots,N_1\}\cup\{N_1+1,\dots,N_1+N_2\}\cup\cdots$. It follows that $\Xvec$ is 1-lumpable w.r.t.\ $g$ and that $\Yvec\sim\Mar{1,\dom{Y},\Amat}$. If $\Amat$ is a permutation matrix, then $\Yvec$ is 1-predictable. Finally, $\Xvec$ is decomposable only if $\Amat=\eye$.

Note that in general a Markov chain described by $\Pvec'$ is not irreducible. However, if $\Emat$ is row stochastic and irreducible, then $\Pvec=(1-\varepsilon)\Pvec'+ \varepsilon \Emat$ is irreducible. In this case, for small $\varepsilon$, $\Xvec$ is called nearly completely decomposable, highly 1-predictable, or quasi-1-lumpable w.r.t.\ $g$.
\end{example}

We finally establish inequalities between the proposed cost functions. From~\cite[Ch.~4.5]{Cover_Information2} follows that, for every $k$, $\Delta_L^k(\Xvec,g)\ge\Delta_L^{k+1}(\Xvec,g)$, and that $\Delta_L^\infty(\Xvec,g):=\lim_{k\to\infty}\Delta_L^k(\Xvec,g)=0$. Similarly, $\Delta_P^k(\Xvec,g)\ge\Delta_P^{k+1}(\Xvec,g)$ and $\Delta_P^\infty(\Xvec,g):=\lim_{k\to\infty}\Delta_P^k(\Xvec,g)=\redrate{\Xvec}-\redrate{\Yvec}\ge 0$. In~\cite[Thm.~1]{GeigerEtAl_OptimalMarkovAggregation} it was shown that $\Delta_D(\Xvec,g)\ge\Delta_L^1(\Xvec,g)\ge\kldr{\Yvec}{\Marx{\Yvec}{1}}$. We can generalize this for every $k$, i.e.,
\begin{equation}
 \Delta_P^k(\Xvec,g)\ge\Delta_L^k(\Xvec,g)\ge\kldr{\Yvec}{\Marx{\Yvec}{k}}.
\end{equation}
Most interestingly, also $\Delta_P^\infty(\Xvec,g)\ge\Delta_L^1(\Xvec,g)$, i.e., the cost functions form an ordered set:
\begin{multline}
 \Delta_D(\Xvec,g)\equiv\Delta_P^1(\Xvec,g) \ge \cdots \ge \Delta_P^\infty(\Xvec,g)\ge\\\Delta_L^1(\Xvec,g)\ge \cdots \ge \Delta_L^\infty(\Xvec,g)=0.
\end{multline}

\subsection{Lumpability, Predictability, and Liftings}\label{sec:Aggregation:lift}

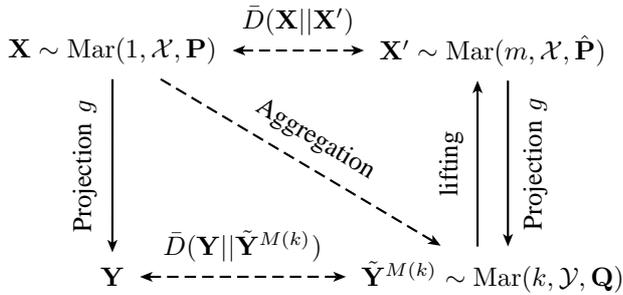
\begin{figure}[t] 
\centering
  \begin{pspicture}[showgrid=false](-1,1)(6,4.75)
    \psset{style=Arrow}
    \pssignal(0,4){x}{$\Xvec\sim\Mar{1,\dom{X},\Pvec}$}
    \pssignal(0,1){y}{$\Yvec$}
    \pssignal(5,1){yp}{$\Marx{\Yvec}{k}\sim\Mar{k,\dom{Y},\Qvec}$}
    \pssignal(5,4){xp}{$\Xvec'\sim\Mar{m,\dom{X},\hat\Pvec}$}
    \ncline{<-}{y}{x} \aput{:U}{Projection $g$}
    \ncline[style=Dash]{x}{yp}\aput{:U}{Aggregation}
    \pnode(4.8,1.4){yp1}\pnode(4.8,3.6){xp1}\pnode(5.2,1.4){yp2}\pnode(5.2,3.6){xp2}
    \ncline{yp1}{xp1}\bput[-15pt]{:U}{lifting}
    \ncline{<-}{yp2}{xp2}\bput[5pt]{:U}{Projection $g$}
    \ncline[style=Dash]{<->}{y}{yp}\Aput{$\kldr{\Yvec}{\Marx{\Yvec}{k}}$}
    \ncline[style=Dash]{<->}{x}{xp}\Aput{$\bar{D}(\Xvec || \Xvec')$}
\end{pspicture}
\caption{Illustration of a lifting: The goal of a lifting is to find an $m$-th order Markov chain $\Xvec'\sim\Mar{m,\dom{X},\hat\Pvec}$ that is $k$-lumpable to the $k$-th order Markov chain $\Marx{\Yvec}{k}$. Such a lifting is always possible for $m=k$.}
\label{fig:liftings}
\end{figure}

Some of the proposed cost functions can be represented via \emph{liftings}. A lifting is a constructive solution to a constrained \emph{realization problem}, i.e., the problem to formulate a stochastic process $\Yvec$ as a projection of a first-order Markov chain $\Xvec$. The unconstrained realization problem can be solved if $\Yvec$ is a Markov chain of any order, since every $\Yvec\sim\Mar{k,\dom{X},\Pvec}$ can be written as a projection of its $k$-transition chain $\Xvec=\blocked{\Yvec}{k}\sim\Mar{1,\dom{X}^k,\tilde\Pvec}$. However, not every stationary process $\Yvec$ on a finite alphabet can be realized as a projection of a Markov chain on a finite alphabet; see~\cite{Anderson_Realization,Vidyasagar_Realization} for an overview of the existing literature and for sufficient and necessary conditions.

In a lifting, illustrated in Fig.~\ref{fig:liftings}, the realization problem is hardened by adding the constraint that the first-order Markov chain $\Xvec$ has a given alphabet, say $\dom{X}$. Liftings of a first-order Markov chain to a first-order Markov chain on a larger alphabet are always possible. In~\cite{Meyn_MarkovAggregation}, the authors lifted $\Marx{\Yvec}{1}$, a first-order Markov chain on $\dom{Y}$, to a first-order Markov chain $\Xvec'\sim\Mar{1,\dom{X},\hat\Pvec}$. Since they employed the invariant distribution vector $\muvec$ of the original chain $\Xvec$, their lifting is called $\muvec$-lifting, and they showed that $\kldr{\Xvec}{\Xvec'}=\Delta_P^1(\Xvec,g)$. Similarly, the authors of~\cite{GeigerEtAl_OptimalMarkovAggregation} suggested a lifting employing the transition matrix $\Pvec$ of the original chain $\Xvec$ to obtain a Markov chain $\Xvec''\sim\Mar{1,\dom{X},\hat\Pvec'}$. For this so-called $\Pvec$-lifting, one gets $\kldr{\Xvec}{\Xvec''}=\Delta_L^1(\Xvec,g)$.

Employing necessary conditions from~\cite{Anderson_Realization}, one can show that it is not always possible to lift the $k$-th order Markov chain $\Marx{\Yvec}{k}$ on $\dom{Y}$ to a first-order Markov chain on $\dom{X}$. However, it is always possible to lift $\Marx{\Yvec}{k}$ to a $k$-th order Markov chain on $\dom{X}$. In particular, the $\muvec$-lifting of $\Marx{\Yvec}{k}\sim\Mar{k,\dom{Y},\Qvec}$ defines $\Xvec'\sim\Mar{k,\dom{X},\hat\Pvec}$, where
\begin{equation}
 \hat{P}_{i_1,i_2,\dots,i_k\to j} = \frac{\mu_j}{\sum_{\ell\in\preim{g(j)}} \mu_\ell}Q_{g(i_1),g(i_2),\dots,g(i_k)\to g(j)}.
\end{equation}
Since the original Markov chain $\Xvec\sim\Mar{1,\dom{X},\Pvec}$ is a trivial $k$-th order Markov chain, we have $\Xvec\equiv\Marx{\Xvec}{k}\sim\Mar{k,\dom{X},\tilde\Pvec}$, where $\tilde{P}_{i_1,i_2,\dots,i_k\to j}=P_{i_k\to j}$. Combining this with~\eqref{eq:KLDRMarkov} yields $\kldr{\Marx{\Xvec}{k}}{\Xvec'}=\Delta_P^k(\Xvec,g)$. To the best of the authors' knowledge, it is not possible to write $\Delta_L^k(\Xvec,g)$, $k>1$, as a KLDR between $\Marx{\Xvec}{k}$ and a lifting of $\Marx{\Yvec}{k}$; in particular, the generalization of $\Pvec$-lifting to higher orders yields an $\Xvec''\sim\Mar{k,\dom{X},\hat\Pvec'}$ such that $\kldr{\Marx{\Xvec}{k}}{\Xvec''}\ge\Delta_L^k(\Xvec,g)$.

\begin{figure*}[t]
 \centering
\subfigure[Cluster error probabilities (CEPs)]{%
\footnotesize
 \begin{tikzpicture}
\begin{axis}[width=0.31\textwidth,height=0.19\textheight,
xmin = 0.1,ymin = 0,%
xmax = 0.9,ymax = 100,%
xlabel={$\varepsilon$},%
ylabel={\%},%
grid=both,%
xlabel near ticks,%
ylabel near ticks,%
legend columns=1,%
legend style={at={(0.01,0.99)},anchor=north west,draw=none},%
legend entries = {{$\mathsf{CEP}_1$},{$\mathsf{CEP}_2$}},%
]%
\addplot[black,no markers,very thick,solid] table {Error_MI};
\addplot[red,no markers,very thick,solid] table {Error_BS};
\end{axis}
\end{tikzpicture}%
}\hfill
\subfigure[$\varepsilon=0.3$; $\mathsf{CEP}_1=19\%$, $\mathsf{CEP}_2=10.2\%$]{\includegraphics[width=0.20\textwidth]{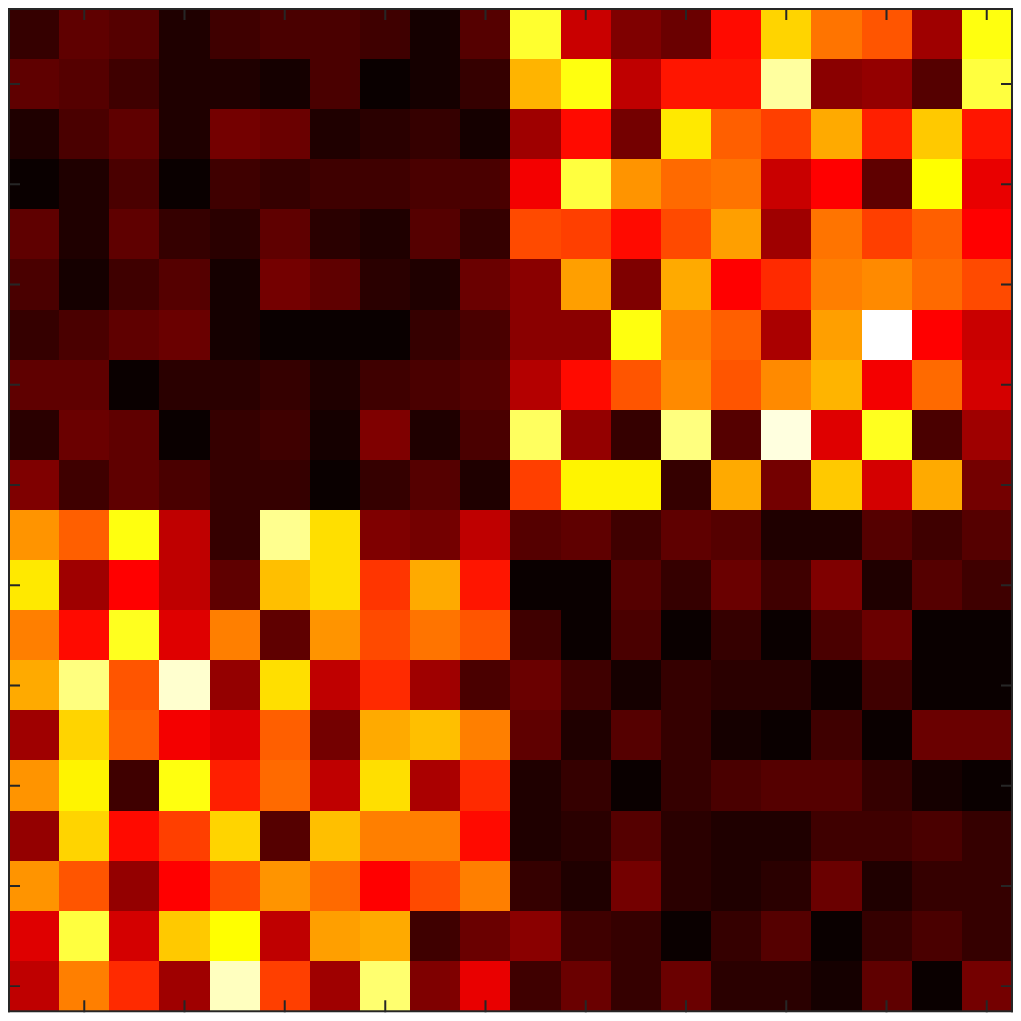}}\hfill
\subfigure[$\varepsilon=0.5$; $\mathsf{CEP}_1=22.4\%$, $\mathsf{CEP}_2=17.6\%$]{\includegraphics[width=0.20\textwidth]{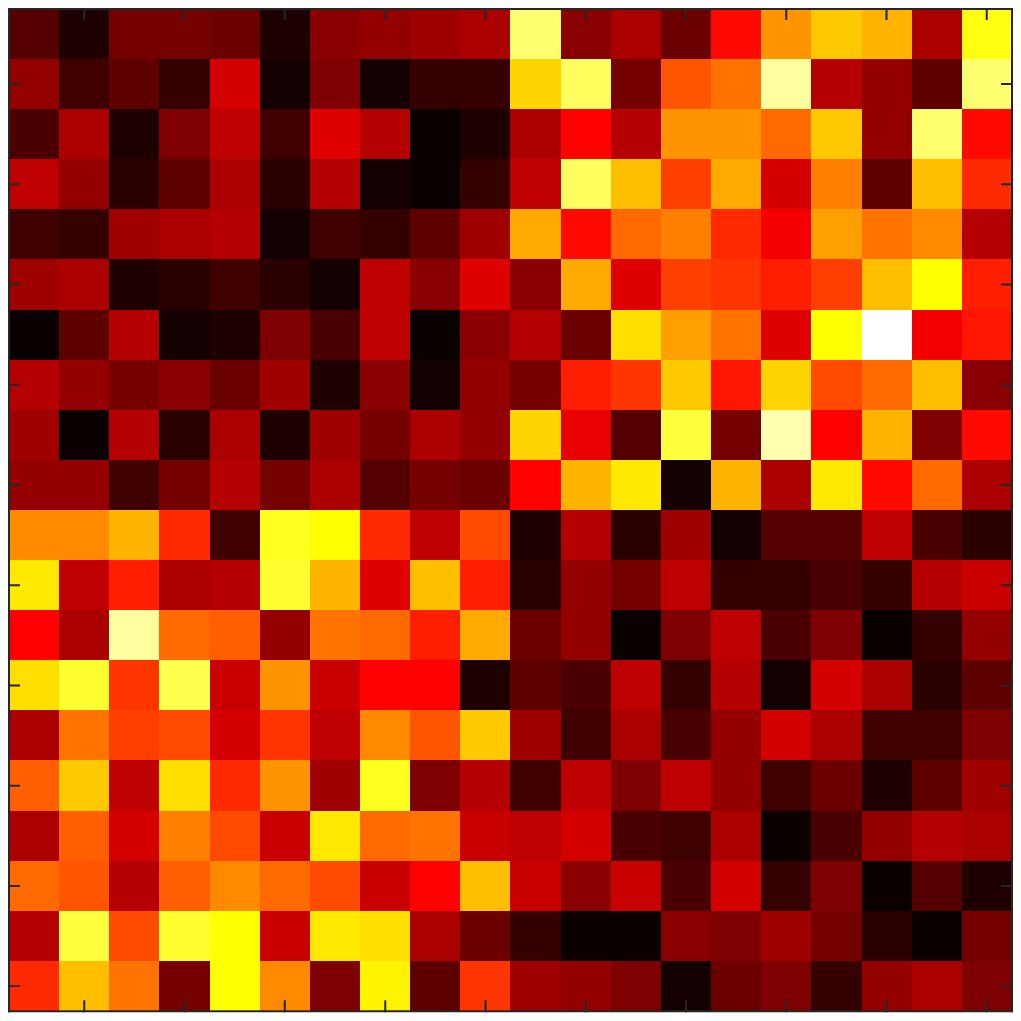}}\hfill
\subfigure[$\varepsilon=0.7$; $\mathsf{CEP}_1=29\%$, $\mathsf{CEP}_2=26.2\%$]{\includegraphics[width=0.20\textwidth]{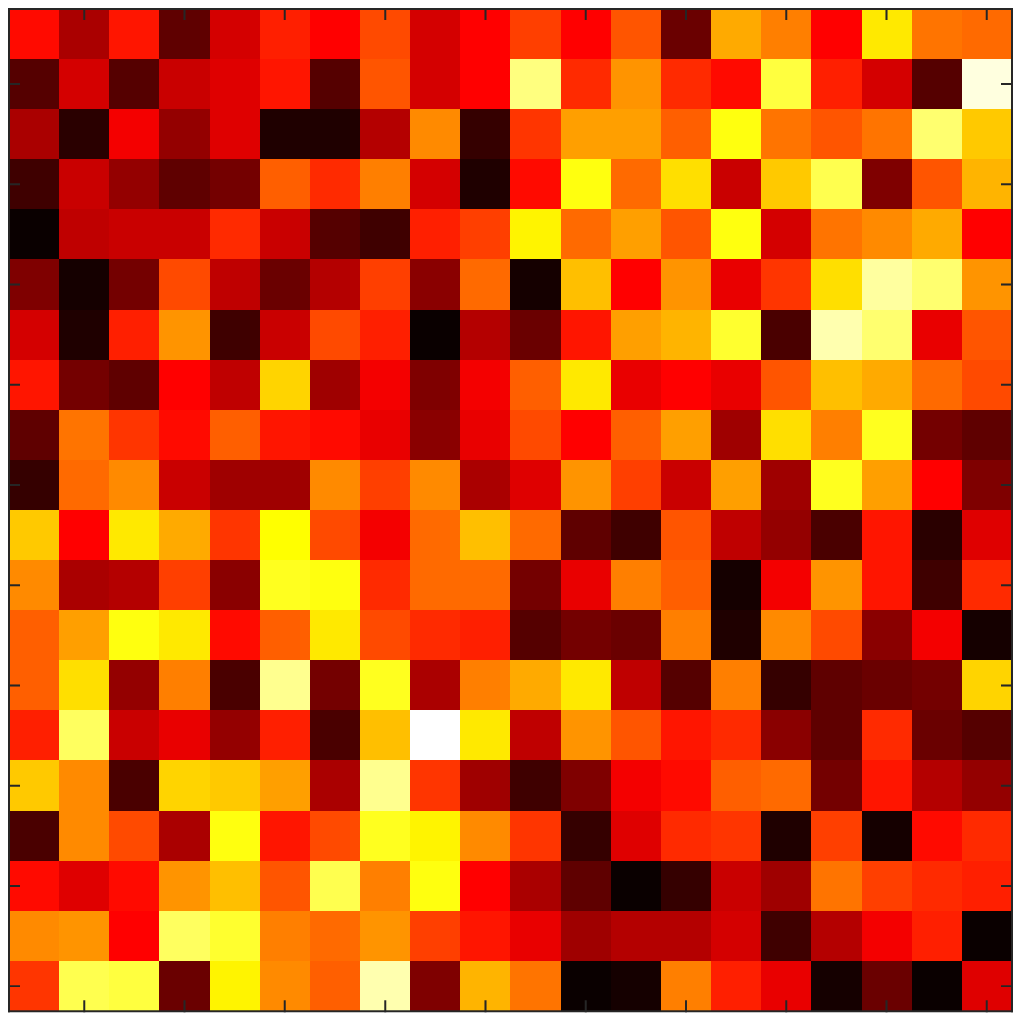}}
\caption{(a) Cluster error probabilities (CEPs) for the quasi-periodic model in Section~\ref{ssec:ex:nearly} as a function of the perturbation parameter $\varepsilon$. A cluster error occurs if at least one state of the Markov chain is misclassified. The results show that $\mathsf{CEP}_2$ obtained by minimizing $\Delta_P^2(\Xvec,g)$ is smaller than $\mathsf{CEP}_1$ obtained by minimizing $\Delta_P^1(\Xvec,g)$. (b)-(d) Colorplots of the matrix $\Pvec=(1-\varepsilon)\Pvec' + \varepsilon\Emat$, for $\varepsilon=0.3$ (b), $\varepsilon=0.5$ (c), and $\varepsilon=0.7$ (d).}
\label{fig:results}
\end{figure*}

\subsection{Aggregation Algorithms}\label{sec:algos}\label{sec:Aggregation:algos}
The Markov aggregation problem, as introduced in Definition~\ref{def:problem}, requires finding a partition of $\dom{X}$ into $M$ non-empty sets. The number of possible such partitions is given by the Stirling number of the second kind, which for large $N$ and fixed $M$ behaves as $M^N/N!$~\cite[24.1.4]{Abramowitz_Handbook}. The problem is thus combinatorial, and sub-optimal algorithms are necessary to attack it in practice. 

One approach is to use \emph{agglomerative} algorithms that rely on merging those two elements of a partition that cause the least cost. These greedy methods usually start with the trivial partition $\dom{X}$ and merge states until the partition has exactly $M$ elements. It follows that these methods benefit from a fixed number of iterations; the fact that elements of the partition are merged furthermore suggests that the computation of the cost function can be simplified in some cases. For example, in~\cite{Alush_PairwiseClustering} the authors employed $\Delta_P^1(\Xvec,g)$ for clustering and showed that the computational complexity of computing the cost can be reduced by using a merging procedure. We believe that similar results hold for $\Delta_P^k(\Xvec,g)$.

A second heuristic are \emph{sequential} algorithms that start with a random partition into $M$ elements. These algorithms proceed by iteratively moving states from their current group into a group that minimizes the cost function, either for a fixed number of iterations over all states or until a local optimum is reached. Since moving states can be interpreted as removing a state from its current group followed by a merging procedure, also here the computation of the cost function can be simplified in some cases, cf.~\cite[Thm.~1]{Alush_PairwiseClustering}. While, depending on the implementation, these sequential heuristics do not guarantee a fixed number of iterations, they can be restarted with different random initializations to avoid local minima.

The authors of~\cite{Meyn_MarkovAggregation} showed that, given that the additive reversibilization of $\Pvec$ satisfies certain properties, at least for $M=2$ the cost function $\Delta_P^1(\Xvec,g)$ is minimized by choosing $g$ according to the sign structure of a specific eigenvector. For general $M$, they suggest recursively splitting groups of states until the desired cardinality of $\dom{Y}$ is reached.

Since $\Delta_L^k(\Xvec,g)$ is not monotonous w.r.t.\ the cardinality $M$ of $\dom{Y}$, agglomerative (or other hierarchical) methods cannot be used effectively. Moreover, preliminary investigations suggest that these cost functions suffer from many local optima, requiring several restarts of the sequential heuristics mentioned above. In~\cite{GeigerEtAl_OptimalMarkovAggregation}, the authors relaxed $\Delta_L^1(\Xvec,g)$ to a cost function that is more well-behaved and that allows employing the information bottleneck method~\cite{Tishby_InformationBottleneck}. We believe that also $\Delta_L^k(\Xvec,g)$ can be relaxed to allow applying either the information bottleneck method or its conditional version~\cite{Gondek_Conditional}.

\section{Examples \& Applications}\label{sec:Applications}
We illustrate our theoretical results at the hand of a few examples.

\subsection{Synthetic Toy Example}\label{ssec:ex:pred}
Consider a first-order Markov chain $\Xvec$ with transition matrix
\begin{equation}
 \Pvec = (1-\varepsilon) \left[ \begin{array}{cccccc}
                 0 & 0 & 1 & 0 & 0 & 0\\
                 0 & 0 & 0 & 1 & 0 & 0\\
                 0 & 0 & 0 & 0 & 1 & 0\\
                 0 & 0 & 0 & 0 & 1 & 0\\
                 p & 1-p & 0 & 0 & 0 & 0\\
                 0 & 0 & 0 & 0 & 0 & 1
                \end{array}
\right]+\varepsilon\Emat
\end{equation}
where $\varepsilon\ll1$ and where $\Emat$ is irreducible and row stochastic. Apparently, this Markov chain is neither nearly completely decomposable nor highly $k$-predictable, for any $k$. However, it can be shown that this Markov chain is quasi-1-lumpable w.r.t.\ the partition $\{\{1,2\},\{3,4\},\{5\},\{6\}\}$. Moreover, it is highly 2-predictable w.r.t.\ the partition $\{\{1,2\},\{3,4\},\{5,6\}\}$: If $X_{n-2}\in\{3,4\}$ and $X_{n-1}\in\{5,6\}$, then with high probability $X_n\in\{1,2\}$; if $X_{n-2}\in\{5,6\}$ and $X_{n-1}\in\{5,6\}$, then with high probability $X_n\in\{5,6\}$.


\subsection{Quasi-Periodic Markov Chain}\label{ssec:ex:nearly}
Let, similar to Example~\ref{ex:blockstochastic},
\begin{equation}
 \Pvec' = \left[ \begin{array}{cc}
                 0 & \Pvec_{12} \\ \Pvec_{21} & 0 
                \end{array}\right],
\end{equation}
where the $10\times 10$ matrices $\Pvec_{ij}$ are row stochastic; hence, $N=20$ and $M=2$. Let $\Emat$ be irreducible, aperiodic, and row-stochastic. Hence, with $\Pvec=(1-\varepsilon) \Pvec'+\varepsilon \Emat$, $\Xvec\sim\Mar{1,\dom{X},\Pvec}$ is aperiodic. We varied $\varepsilon$ from 0.1 to 0.9 in steps of 0.1. Note that several eigenvalues of the additive reversibilization of $\Pvec$ are negative. Thus, Assumption~2 in~\cite{Meyn_MarkovAggregation} is violated and the proposed eigenvector-based aggregation algorithm cannot be applied to minimize $\Delta_P^1(\Xvec,g)$.

Choosing $g$ according to the natural partition of $\Pvec$ makes $\Yvec$ nearly periodic and highly 1-predictable, hence highly $k$-predictable for every $k$. We generated 500 random\footnote{Row stochastic matrices were generated by choosing entries uniformly from $[0,1]$ normalizing the rows subsequently.} matrices $\Pvec$ and applied a sequential algorithm similar to the one in~\cite[Tab.~1]{Alush_PairwiseClustering} to minimize $\Delta_P^1(\Xvec,g)$ and $\Delta_P^2(\Xvec,g)$. The results in Fig.~\ref{fig:results} show that minimizing $\Delta_P^2(\Xvec,g)$ leads to a higher probability of detecting the natural partition, thus trading accuracy for computational complexity. Preliminary analyses suggest that this can be explained by the minimization of $\Delta_P^2(\Xvec,g)$ getting less likely stuck in local minima.

\subsection{Aggregating a Letter Bi-Gram Model}

\begin{table}[t]
 \caption{Aggregating a Letter Bi-Gram Model}
 \label{tab:bigram}
 \centering
 \begin{tabular}{l|l}
  Cost & Partition of the alphabet\\
  \hline
  $\Delta_P^1(\Xvec,g)$ & \texttt{\textvisiblespace'Zx}\\
 	& \texttt{!\$(1?BCDFGHJKLMNPQRSTVWY[bchjmpqvw}\\
 	& \texttt{"23456789EUaeiou\'{e}}\\
	& \texttt{),-.0:;AIO]dfgklnrsty}  \\
  \hline
  $\Delta_P^2(\Xvec,g)$ & \texttt{\textvisiblespace!'),.03:;?]}\\
	      & \texttt{bcdfgklmnprstvwxy}\\
	      & \texttt{aeiou\'{e}}\\
	      & \texttt{"\$(-12456789ABCDEFGHIJKLMNOPQRSTUVWYZ[hjq}
 \end{tabular}

\end{table}

We trained a letter bi-gram model of F. Scott Fitzgerald's book ``The Great Gatsby''. We modified the text by removing all chapter headings and line breaks, but kept punctuation unchanged. The letter bi-gram model is a first-order Markov chain on an alphabet of size $N=76$ (upper and lower case letters, numbers, punctuation, etc.) and was trained from the remaining 266615 characters of the text. We applied a sequential algorithm for $M=4$ ten times with random initializations to minimize $\Delta_P^1(\Xvec,g)$ and $\Delta_P^2(\Xvec,g)$. The best results for either cost function are displayed in Table~\ref{tab:bigram}.

It can be seen that minimizing $\Delta_P^2(\Xvec,g)$ leads to a more meaningful partition of the alphabet, except for a few misclassified states: lower case vowels, lower case consonants, space and punctuation usually followed by a space, and numbers, upper case letters, and punctuation usually preceded by a space. The partition obtained by minimizing $\Delta_P^1(\Xvec,g)$ does not appear to be meaningful. We therefore believe that the proposed extension to higher-order aggregations admits linguistic interpretations and might be useful for the reduction of models in natural language processing. 

\subsection{Simplifying a Maintenance Model}
\begin{figure}[t]
\centering
  \begin{tikzpicture}[-latex ,auto ,node distance =1.75 cm and 1.75 cm ,on grid ,thick ,state/.style ={ circle ,top color =white , bottom color = black!20 ,draw,black , text=black , minimum width =0.5 cm},rounded corners]
\tikzset{>=latex}
\footnotesize
\node[state] (W) {$W$};
\node[state] (D1) [right =of W] {$D_1$};\node[state] (M1) [below =of D1] {$M_1$};
\node[state] (D2) [right =of D1] {$D_2$};\node[state] (M2) [below =of D2] {$M_2$};
\node[state] (D3) [right =of D2] {$D_3$};\node[state] (M3) [below =of D3] {$M_3$};
\node[state] (F1) [right =of D3] {$F_1$};\node[state] (M4) [below =of F1] {$M_4$};
\node[state] (F0) [above = 2cm of D2] {$F_0$};

\path (W) edge[bend right=45, ->] node[above,inner sep=0pt,fill=white]{$\lambda_m$} (M1);
\path (W) edge[->] node{$\lambda_1$} (D1);
\path (D1) edge[->] node{$\lambda_1$} (D2);
\path (D2) edge[->] node{$\lambda_1$} (D3);
\path (D3) edge[->] node{$\lambda_1$} (F1);
\path (D1) edge[->] node[above,inner sep=0pt,fill=white]{$\lambda_m$} (M2);
\path (D2) edge[->] node[above,inner sep=0pt,fill=white]{$\lambda_m$} (M3);
\path (D3) edge[->] node[above,inner sep=0pt,fill=white]{$\lambda_m$} (M4);

\path (W) edge[->] node[midway,inner sep=1pt]{$\lambda_0$} (F0);
\path (D1) edge[->] node[pos=0.5,inner sep=0pt,fill=white]{$\lambda_0$} (F0);
\path (D2) edge[->] node[pos=0.5,inner sep=0pt,fill=white]{$\lambda_0$} (F0);
\path (D3) edge[->] node[pos=0.5,inner sep=0pt,fill=white]{$\lambda_0$} (F0);

\path (M1) edge[->] node[above,inner sep=0pt,fill=white]{$\mu_m$} (W);
\path (M2) edge[->] node[above,inner sep=0pt,fill=white]{$\mu_m$} (W);
\path (M3) edge[->] node[above,pos=0.4,inner sep=0pt,fill=white]{$\mu_m$} (D1);
\path (M4) edge[->] node[above,inner sep=0pt,fill=white]{$\mu_m$} (D2);

\path (F0) edge[bend right,->] node[above]{$\mu_0$} (W);
\path (F1) edge[bend right=90,->] node[pos=0.75,above]{$\mu_1$} (W);

\node[draw,inner sep=2mm,label=left:\textcolor{red}{1},fit=(W),draw=red] {};
\node[draw,inner sep=2mm,label=right:\textcolor{red}{6},fit=(F0),draw=red] {};
\node[draw,inner sep=2mm,label=below:\textcolor{red}{2},fit=(M1) (D1),draw=red] {};
\node[draw,inner sep=2mm,label=below:\textcolor{red}{3},fit=(M2) (D2),draw=red] {};
\node[draw,inner sep=2mm,label=below:\textcolor{red}{4},fit=(M3) (D3),draw=red] {};
\node[draw,inner sep=2mm,label=below:\textcolor{red}{5},fit=(F1) (M4),draw=red] {};
\end{tikzpicture}
\caption{Maintenance model from~\cite{Chan_Maintenance} for $k=4$. $W$ indicates the working state, $D_1$ through $D_3$ indicate various states of deterioration, and $M_1$ through $M_4$ are maintenance states. A system may fail through deterioration ($F_1$) or spontaneously ($F_0$); in both cases, the system is repaired and brought back to working condition. The edge labels indicate the rates with which these transitions occur in a continuous-time model. The red boxes indicate a partition that makes the resulting process $\Yvec$ at least partly 2-predictable.}
\label{fig:maintenance}
\end{figure}
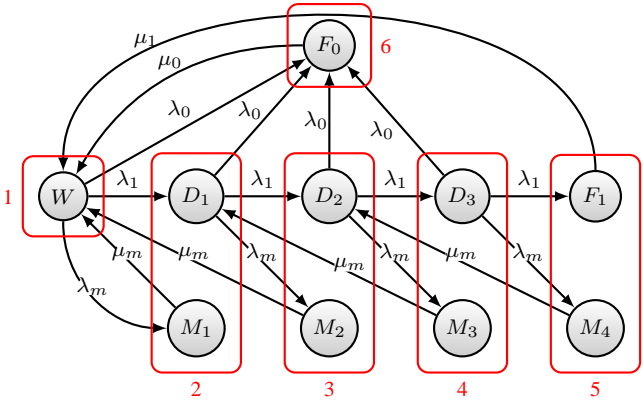

We finally investigate the continuous-time maintenance model with deterioration discussed in~\cite{Chan_Maintenance}. In this model, the system fails either spontaneously or after $k$ steps of deterioration. The system furthermore undergoes maintenance after random intervals, revoking the last deterioration step. After failure, the system is repaired to full working condition. This model, a Markov chain with $N=2k+4$ states, is depicted in Fig.~\ref{fig:maintenance}, where the edge labels indicate the rates with which transitions occur. We generated the embedded discrete-time Markov chain $\Xvec$ by converting the rate matrix to a transition probability matrix. It can be shown that the aggregation to $M=k+3$ states indicated in Fig.~\ref{fig:maintenance} admits predicting $Y_n$ from two past states $Y_{n-1}$ and $Y_{n-2}$: If, for $\ell\in\{2,\dots,k+1\}$, we have $Y_{n-2}=\ell+1$ and $Y_{n-1}=\ell-1$, $X_{n-1}$ is in a (deteriorated) working state, and it is very likely that $Y_n=\ell$. We applied a sequential algorithm minimizing $\Delta_P^2(\Xvec,g)$ to this model, for $M=k+3$ and $k$ ranging between 3 and 7. The algorithm found the indicated partition, where for larger values of $k$ the results were more reliable if the maintenance rate $\lambda_m$ was smaller than the deterioration rate $\lambda_1$.

Prediction from only one past state fails in this model: From $Y_{n-1}=\ell$, both $Y_{n}=\ell+1$ and $Y_{n}=\ell-1$ are possible, for $\ell\in\{2,\dots,k-1\}$. Thus, applying our sequential algorithm to minimize $\Delta_P^1(\Xvec,g)$ did not produce meaningful results.

\section*{Acknowledgments}
The work of Bernhard C. Geiger was funded by the Erwin Schr\"odinger Fellowship J 3765 of the Austrian Science Fund.

\bibliographystyle{IEEEtran}
\bibliography{IEEEabrv,../../References/InformationProcessing,%
../../References/ProbabilityPapers,%
../../References/textbooks,%
../../References/myOwn,%
../../References/UWB,%
../../References/InformationWaves,%
../../References/ITBasics,%
../../References/SignalProcessing,%
../../References/HMMRate,%
../../References/MarkovStuff,%
../../References/ITAlgos}

\end{document}

%% file: abbrevations_processing.tex
\newcommand{\ent}[1]{H(#1)}

\newcommand{\mutinf}[1]{I(#1)}

\newcommand{\kldrate}[2]{\bar{D}(\mathbf{#1}||\mathbf{#2})}
\newcommand{\kldr}[2]{\bar{D}({#1}||{#2})}
\newcommand{\binent}[1]{H_2(#1)}

\newcommand{\entrate}[1]{\bar{H}(\mathbf{#1})}

\newcommand{\redrate}[1]{\bar{R}(\mathbf{#1})}

\newcommand{\loss}[2][\empty]{\ifthenelse{\equal{#1}{\empty}}{L(#2)}{L_{#1}(#2)}}
\newcommand{\lossrate}[2][\empty]{\ifthenelse{\equal{#1}{\empty}}{L(\mathbf{#2})}{L_{\mathbf{#1}}(\mathbf{#2})}}

\newcommand{\relLoss}[2][\empty]{\ifthenelse{\equal{#1}{\empty}}{l(#2)}{l_{#1}(#2)}}

\newcommand{\dom}[1]{\mathcal{#1}}

\newcommand{\Prob}[1]{\mathrm{Pr}(#1)}
\newcommand{\Mar}[1]{\mathrm{Mar}(#1)}

\newcommand{\Xvec}{\mathbf{X}}

\newcommand{\Yvec}{\mathbf{Y}}

\newcommand{\Zvec}{\mathbf{Z}}

\newcommand{\Pvec}{\mathbf{P}}
\newcommand{\muvec}{\boldsymbol{\mu}}

\newcommand{\Wvec}{\mathbf{W}}

\newcommand{\Amat}{\mathbf{A}}

\newcommand{\eye}{\mathbf{I}}

\newcommand{\perr}{P_e}

\newcommand{\preim}[1]{g^{-1}(#1)}

\newcommand{\limn}{\lim_{n\to\infty}}

